\documentclass[fleqn,usenatbib]{mnras}

\usepackage{newtxtext,newtxmath}

\usepackage[T1]{fontenc}

\DeclareRobustCommand{\VAN}[3]{#2}
\let\VANthebibliography\thebibliography
\def\thebibliography{\DeclareRobustCommand{\VAN}[3]{##3}\VANthebibliography}

\usepackage{graphicx}	
\usepackage{amsmath}	

\title[QPOs of MAXI J1803$-$298 ]{Wavelet analysis of low-frequency quasi-periodic oscillations in MAXI J1803$-$298 observed with \textit{Insight-HXMT} and \textit{NICER}}
\author[Jin et al.]{
Y. J. Jin,$^{1,2}$
X. Chen,$^{2}$
H. F. Zhu,$^{2}$
Z. J. Jiang,$^{1}$
L. Zhang,$^{1}$
W. Wang$^{2}$\thanks{wangwei2017@whu.edu.cn}
\\
$^{1}$Department of Astronomy, School of Physics and Astronomy, Yunnan University, Kunming 650500, China \\
$^{2}$Department of Astronomy, School of Physics and Technology, Wuhan University, Wuhan 430072, China\\
}

\date{Accepted XXX. Received YYY; in original form ZZZ}

\pubyear{2024}

\begin{document}
\label{firstpage}
\pagerange{\pageref{firstpage}--\pageref{lastpage}}
\maketitle

\begin{abstract}
With data observed by the Hard X-ray Modulation Telescope (\textit{Insight}-HXMT) and the Neutron star Interior Composition Explorer (\textit {NICER}), we study low-frequency quasi-periodic oscillations (LFQPOs) of the black hole candidate MAXI J1803$-$298 during the 2021 outburst. Based on hardness intensity diagram and difference of the QPOs properties, Type-C and Type-B QPOs are found in the low-hard state and soft intermediate state, respectively. After searching for and classifying QPOs in Fourier domains, we extract the QPO component and study it with wavelet analysis. The QPO and no-QPO time intervals are separated by the confidence level, so that the S-factor, which is defined as the ratio of the QPO time interval to the total length of good time interval, is calculated. We found S-factors decrease with QPOs frequency for Type-C QPOs but stay stable around zero for Type-B QPOs. The relation of S-factor of Type-C QPOs and photon energy, the correlation of S-factor and counts are also studied. Different correlation of S-factor and counts for different energy bands indicates different origins of QPOs in high energy and low energy bands, which may be due to a dual-corona scenario.
\end{abstract}

\begin{keywords}
accretion: accretion discs -- black hole physics -- X-rays: binaries -- stars: individual: MAXI J1803$-$298
\end{keywords}



\section{Introduction}
\label{sec:in}

Black hole X-ray binary (BHXRB) is the system which consist of a black hole and a companion star. Most of BHXRBs are usually in quiescence and sometimes in outburst last for a few months, which are called the transient systems. During the outburst, vast amounts of energy are released into space, especially in X-rays, making the systems bright enough to be observed \citep{Remillard2006}. \cite{Belloni2005} describe one BHXRBs outburst with several accretion states: the low-hard state (LHS), the hard-intermediate state (HIMS), the soft-intermediate state (SIMS) and the high-soft state (HSS). However, it's important to note that some BHXRBs outbursts indeed show all these four accretion states, and some may show only certain accretion states. The accretion states of BHXRBs outburst can be tracked with the Q-shaped loop in the hardness intensity diagram (HID) diagram.

Low-frequency quasi-periodic oscillations in black hole X-ray binaries were reported for the first time in 1983 by \cite{Motch1983}. Since then, LFQPOs have been studied as a tool to explore the strong gravitational fields and motion of accretion flow around black holes, but it is always a challenge to figure out its physical origin\citep{Stella1998,Jeremy2006,Ingram2009,Cabanac2010,Karpouzas2020,García2021}. LFQPOs are normally found in the power density spectrum (PDS) and classified into Type-A, -B, -C QPOs, based on the following properties: the centroid frequency, the quality factor (Q=$\nu_{\rm centroid}/{\rm FWHM}$) , the fractional root-mean-square (rms) variability and the broad-band noise\citep{Remillard2002,Motta2011,Motta2015,Ingarm2019}.

Type-C QPOs are the most common type of QPOs, which are characterized with high rms, high quality factor, wide frequency range($\sim$0.1 to $\sim$30 Hz) and the Flat-Top Noise (FTN) at a frequency comparable to the QPO frequency \citep{Casella2005}. Type-C QPOs can be observed in basically all states of outburst, but mostly in the LHS and HIMS \citep{Ingarm2019}. In recent researches, geometric models, in particular the relativistic precession models, are preferred to be the origin of Type-C QPOs \citep{Ingram2009,Ingram2010,Ingram2011,Veledina2013,Ingram2015,Motta2015,Ingram2016,Eijnden2017}.
Type-B QPOs only appear in the SIMS, with low rms, $\sim$6 Hz centroid frequency and the weak red noise at low frequency \citep{Wijnands1999}. Type-B QPOs are considered as one of the characteristic features that distinguishes the SIMS from HIMS \citep{Belloni2016}. There is evidence to suggest that Type-B QPOs may be related to the jets \citep{Fender2004,Corbel2005,Stevens2016,Ruiter2019}. Type-A QPOs are very rare and basically appear in the soft state. However, some studies suggest that some properties of Type-A QPOs are similar to that of Type-C QPOs \citep{Motta2012}.

Most of the studies of LFQPOs are completed with Fourier transform \citep{Miyamoto1991,Zhang2017,Huang2018}. LFQPOs can be accurately and efficiently found in Fourier domain and their parameters can be obtained in the power density spectrum (PDS). However, Fourier transform has a limit on studying the rapid changes of LFQPOs in the time domain, due to the tens of seconds time intervals of PDSs \citep{Zhang2021,Sriram2021}. Luckily, with wavelet transform, we can obtain the time–frequency spectrum of irregular and non-continuous signals, allowing us to study the rapid changes of LFQPOs with time intervals as short as seconds\citep{Ding2017,Chen2022a}.

Wavelet analysis has been widely used to study QPOs in many researches\citep{Czerny2010,Ding2017,Tian2023,Ren2023}, including the study of LFQPOs \citep{Lachowicz2010,Chen2022b}. With wavelet analysis, \cite{Chen2022a} found the time intervals can be separated to the time intervals with LFQPOs and without LFQPOs, due to the rapid changing behaviors of LFQPOs appearing and disappearing in a short time interval. With these two time intervals, the PDSs and energy spectra obtained are different. On this basis, \cite{Chen2023} extract LFQPOs component and study them in wavelet analysis. Extracting LFQPOs can effectively reduce the influence of noise and improve the accuracy of separating the time intervals with and without LFQPOs. A significant difference in the S-factor between Type-C QPO and Type-B QPO was found, which may be used for LFQPO classification.

MAXI J1803$-$298 was firstly detected on 2021 May 1 by the Monitor of All-sky X-ray Image (\textit {MAXI}; \cite{Matsuoka2009}) and then observed by other telescopes during its 2021 outburst \citep{Jana2022,Shidatsu2022,Coughenour2023,Wood2023,Zhu2023}. The mass function of the binary was estimated as f(M) = $ 2.1 - 7.2 M_{\sun}$ with the periodicity of 7.02 $\pm 0.18 h$ detected in the light curve\citep{Jana2022}. Based on the X-ray spectroscopy, the mass and spin of the black hole were estimated to be $ 8.5 - 16 M_{\sun}$ and 0.7 \citep{Chand2022}. However, \cite{Feng2022} gave a higher spin about 0.991 and the high inclination about 70 degree for MAXI J1803$-$298 with the reflection model.  Type-C QPOs with frequency from 5.31 to 7.61 Hz were detected during this outburst with AstroSat \citep{Jana2022}. \cite{Chand2022} found that the Type-C QPO frequency and fractional rms amplitude in the 15-30 keV band are higher than those in the 3-15 keV band. Type-B QPOs are also detected during this outburst with Insight-HXMT and NICER \citep{Zhu2023}. The lag spectrum of type-B QPOs exhibits a U-shaped pattern which can be explained by the dual-corona model.

In this work, we study the type-C QPOs and type-B QPOs observed in MAXI J1803$-$298 during its 2021 outburst with Insight-HXMT and NICER observations. Except Fourier transform, the wavelet transform is used to study LFQPOs. In Section~\ref{sec:ob}, we describe the observations and methods of data analysis. The QPO studies and wavelet results are presented in Section~\ref{sec:re}. The discussions follow in Section~\ref{sec:di} and we make conclusions in Section~\ref{sec:co}.

\section{Observation and data analysis}
\label{sec:ob}

In this work, we study the QPOs found in the 2021 outburst of MAXI J1803$-$298. The 2021 outburst of MAXI J1803$-$298 was observed by \textit {Insight}-HXMT from May 3 to July 28 with only type-C QPO detected. In addition, \textit{NICER} also observed the source from 2021 May 2 to November 7, and type-B QPOs are found in May \citep{Zhu2023}, thus are included in our study.


For the \textit {Insight}-HXMT data, we use the \textit {Insight}-HXMT Data Analysis software (HXMTDAS) v2.04 and make the light curves with tasks \textit{helcgen}, \textit{melcgen} and \textit{lelcgen}. The time bins of the light curves are set to be 0.0078125s. The good time intervals are generated on the following criteria: the pointing offset angle less than 0.04$^\circ$, the elevation angle greater than 10$^\circ$ and the geomagnetic cutoff rigidity greater than 8$^\circ$. In addition, data within 300 s of the South Atlantic Anomaly (SAA) are excluded. The background of spectra and light curves are obtained with tools (\textit{HEBKGMAP}, \textit{MEBKGMAP} and \textit{LEBKGMAP}).
We choose the  1-10 keV (LE),  10-20 keV(ME) and 28-100 keV(HE) for analysis of \textit {Insight}-HXMT data.
For the \textit {NICER} data, we used the NICERDAS ver.10 software with the calibration files, CALDB xti20221001. The light curves are generated by the tasks nicerl2 and nicerl3-lc of NICER standard pipeline. To compare with the \textit {Insight}-HXMT LE data, only data within the 1-10 keV, which is the same energy band as the \textit {Insight}-HXMT LE data, is used for \textit {NICER} in this work.

The averaged power density spectrum (PDS) is obtained with 64s data intervals and 1/64s time resolution for each observation. The PDS was subjected to fractional rms normalization \citep{Miyamoto1991} and then fitted with several Lorentzian functions using XSPEC v12.12.1. A power-law function is added for fitting the constant white noise. We pick out the Lorentzian component corresponding to the QPO and obtain its parameters (i.e., the centroid frequency $\upsilon_{\rm QPO}$, the FWHM, the QPO rms) for studying their evolution with outburst. We divide integral of the LFQPO by its mean error to show the significance of QPOs \citep{Jin2023}. The total fractional rms variability (the total rms) is also estimated in the range of $0.01-32$ Hz of LE ($1-10$ keV) data.

We further study the time-frequency space of light curves with wavelet analysis based on the wavelet analysis methods introduced by \cite{Torrence1998}. We choose Morlet wavelet with non-dimensional frequency $\omega_0$ = 6 to perform the wavelet transform. The time step and total length of discrete time sequence is set to 0.0078125s (same with the time bin of light curve) and the length of GTI chosen from light curve, respectively. The wavelet transform and the wavelet power are calculated based on the equations in \citep{Torrence1998,Chen2022a}. With the changes of time and frequency, the diffuse two-dimensional time–frequency wavelet power spectrum is obtained (right panel of Figure~\ref{fig:complete_wavelet}). In order to solve the problems caused by dealing with the finite time series, zeros are filled in at the end of the time series temporarily, leading a decrease in amplitude near the edges. Cone of influence (COI, shaded area of Figure~\ref{fig:complete_wavelet}) is the area where the edge effects are severe in the wavelet spectrum. More detailed introduction and equations of wavelet analysis can be found in \cite{Torrence1998}.

To determine significance levels of feature in wavelet analysis, the univariate lag-1 autoregressive [AR(1)] is used for the red-noise background calculation. By multiplying the background spectrum by the chi-square values at a certain confidence level, the corresponding confidence spectrum (the red line in the left panel and the black circled area in the right panel of Figure~\ref{fig:complete_wavelet}) is calculated. If the wavelet power of a signal is above the confidence spectrum, it can be considered as a valid signal in our case. Within a certain frequency range, the sum of the time with wavelet power greater than the confidence spectrum is the effective oscillation time in the corresponding frequency range with corresponding confidence level. In brief, the calculation of the effective oscillation time is affected by the frequency range and the confidence level. When the effective time is calculated in the the frequency range of the QPO FWHM, the effective oscillation time is also called the QPO time interval, while the no-QPO time interval is made up by all the remaining time. Thus, a unique property of QPO based on wavelet analysis, i.e. the ratio of the QPO time interval and the total length of good time interval, called S-factor \citep{Ding2017,Chen2022b}, can be calculated under the corresponding confidence level. Considering that some QPOs may have broad FWHM and multiple S-factors are calculated within the frequency range of the QPO FWHM, we use the average value of multiple S-factors as the S-factors. And the error of the S-factor is calculated based on the formula of standard deviation. In this paper, we mainly study wavelet results with 68\% confidence level, but the results with 95\% confidence level are shown for comparison in Figure~\ref{fig:S_freq}. 

Since reliable results can only be obtained with smooth continuous time intervals \citep{Chen2022b}, we use each separated GTI of light curves for wavelet analysis. In addition, the too short time interval also affect the results, so only the good time intervals (GTIs) longer than 192s (3 times of the time interval of PDS) are used for wavelet analysis and S-factor calculation. Comparing with the whole time of light curve, QPOs in PDSs become relatively weaker with separated GTIs and the noise has a relatively greater impact on the wavelet analysis. Hence, we extract the QPO component, apply it on wavelet analysis and obtain the wavelet result with QPO component extracted. First, we find the Lorentz component corresponding to QPO among several Lorentz components in the PDS. And then we calculate the power ratio of the picked QPO component to the whole PDS with frequency (i.e. the QPO proportion curve). Finally, the QPO proportional curve is multiplied to the wavelet power spectrum, and the wavelet result with QPO component extracted is obtained \citep[see][for the technique details]{Chen2023}. The QPO component extracted wavelet results are shown in Figure~\ref{fig:wavelet}. This method effectively eliminates the influence of noise on the wavelet power spectrum and make the separation of the QPO and no-QPO time intervals more accurately, which means the S-factor is calculated more accurately.

To study the energy dependence of QPOs, we separate all the light curve of \textit {Insight}-HXMT with Type-C QPOs detected into nine energy bands (1-3 keV, 3-5 keV, 5-10 keV, 10-13 keV, 13-16 keV, 16-20 keV, 28-40 keV, 40-60 keV and 60-100 keV). For each GTI (longer than 192s) of each light curve in each energy band, we perform wavelet analysis with 68\% confidence level to obtain a QPO time interval and a no-QPO time interval, so that the S-factor of Type-C QPO can be calculated. By studying the S-factor in different energy bands, we found that the S-factor may have a correlation with counts. Therefore, we calculate the count rate for all the light curves, and study the correlation of counts and S-factor in different energy bands. The linear correlations between S-factor and counts are found in most of the observations. We fit them with linear regression using python package \textit{scikit-learn} and study properties of the linear correlation. In addition, we calculate the Pearson correlation coefficients (PCCs) to measure the linear correlations. In addition, different correlations between S-factor and counts are found in different energy bands.

\section{Results}
\label{sec:re}

\subsection{Light curve}
\label{sec:light} 
\begin{figure}
        \includegraphics[width=\columnwidth]{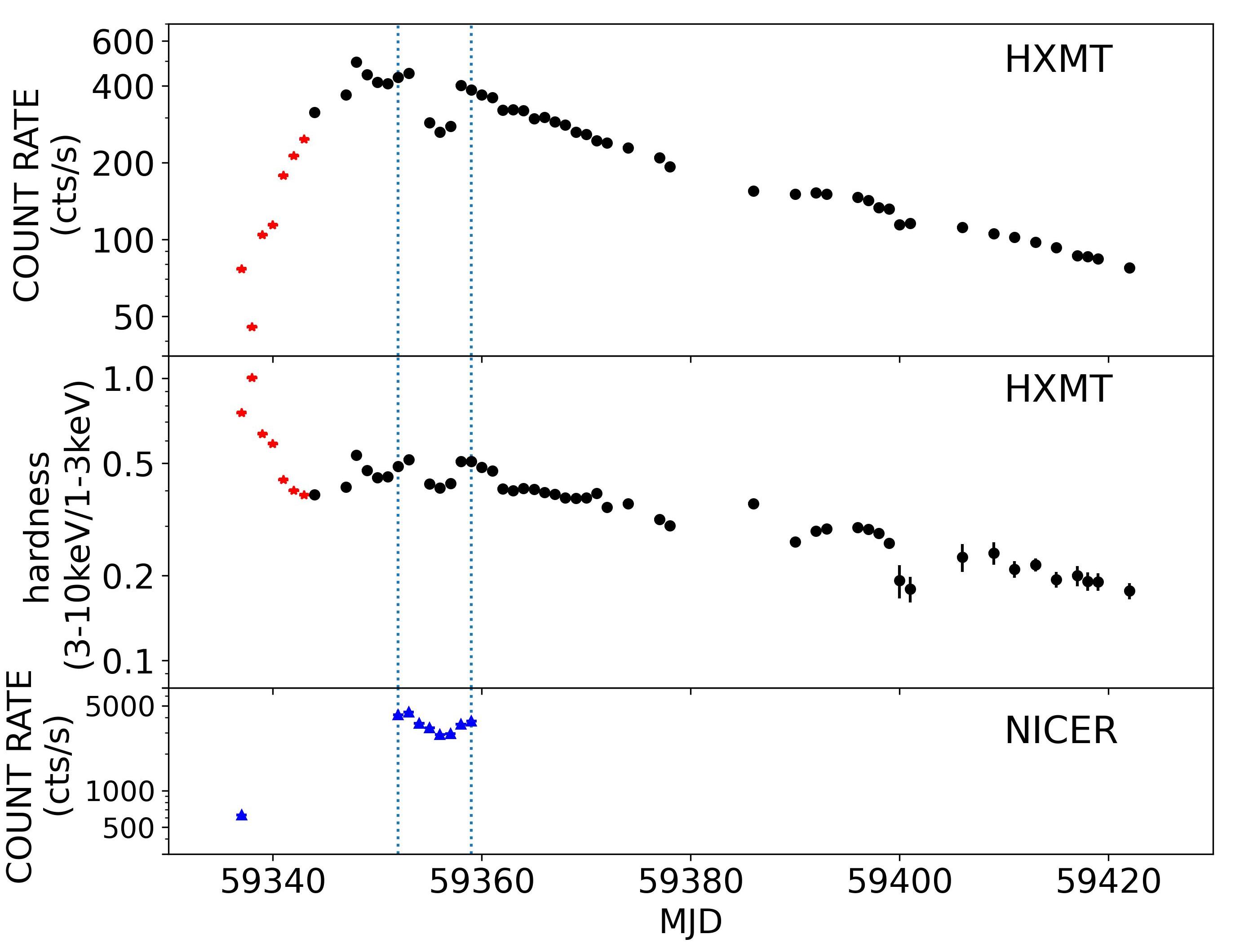}
    \caption{The count rate curves of MAXI J1803$-$298 are presented in the top panel with  \textit {Insight}-HXMT LE ($1-10$ keV) data. Each point represents one day and the red stars represent days with QPOs detected. The uncertainties of count rates and hardness ratio are less than 5\%, then errors bars are smaller than the size of the symbols. The hardness ratios are calculated from the data obtained by  \textit {Insight}-HXMT with ($3-10$ keV) / ($1-3$ keV) and shown in the middle panel. The bottom panel shows the light curve with  \textit {NICER} ($1-10$ keV) data in the same outburst. The dotted line marks the days with QPOs detected by  \textit {NICER} but not by  \textit {Insight}-HXMT.}
    \label{fig:light curve}
\end{figure}

\begin{figure}
        \includegraphics[width=\columnwidth]{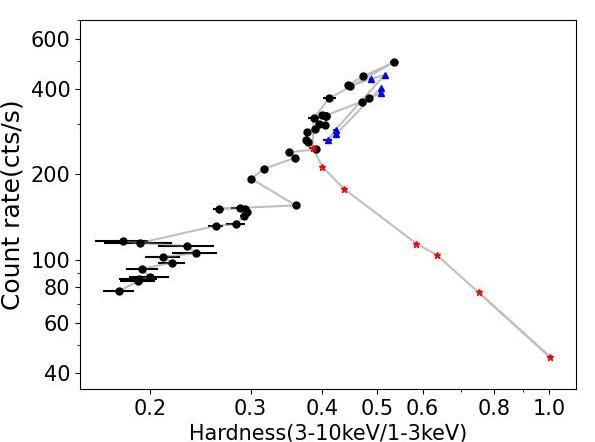}
    \caption{The hardness intensity diagram of MAXI J1803$-$298 during the 2021 outburst. The uncertainties of count rates and hardness ratio are less than 5\%, then errors bars are smaller than the size of the symbols. The Y-axis represents the count rate of LE from $1-10$ keV. The hardness ratio is defined as the count ratio between the energy bands $3-10$ keV and $1-3$ keV. Each point represents one day. The red stars represent the days with QPO detected by  \textit {Insight}-HXMT and the blue triangles represent the days with QPO detected by  \textit {NICER}. }
    \label{fig:HID}
\end{figure}
\begin{figure}
        \includegraphics[width=\columnwidth]{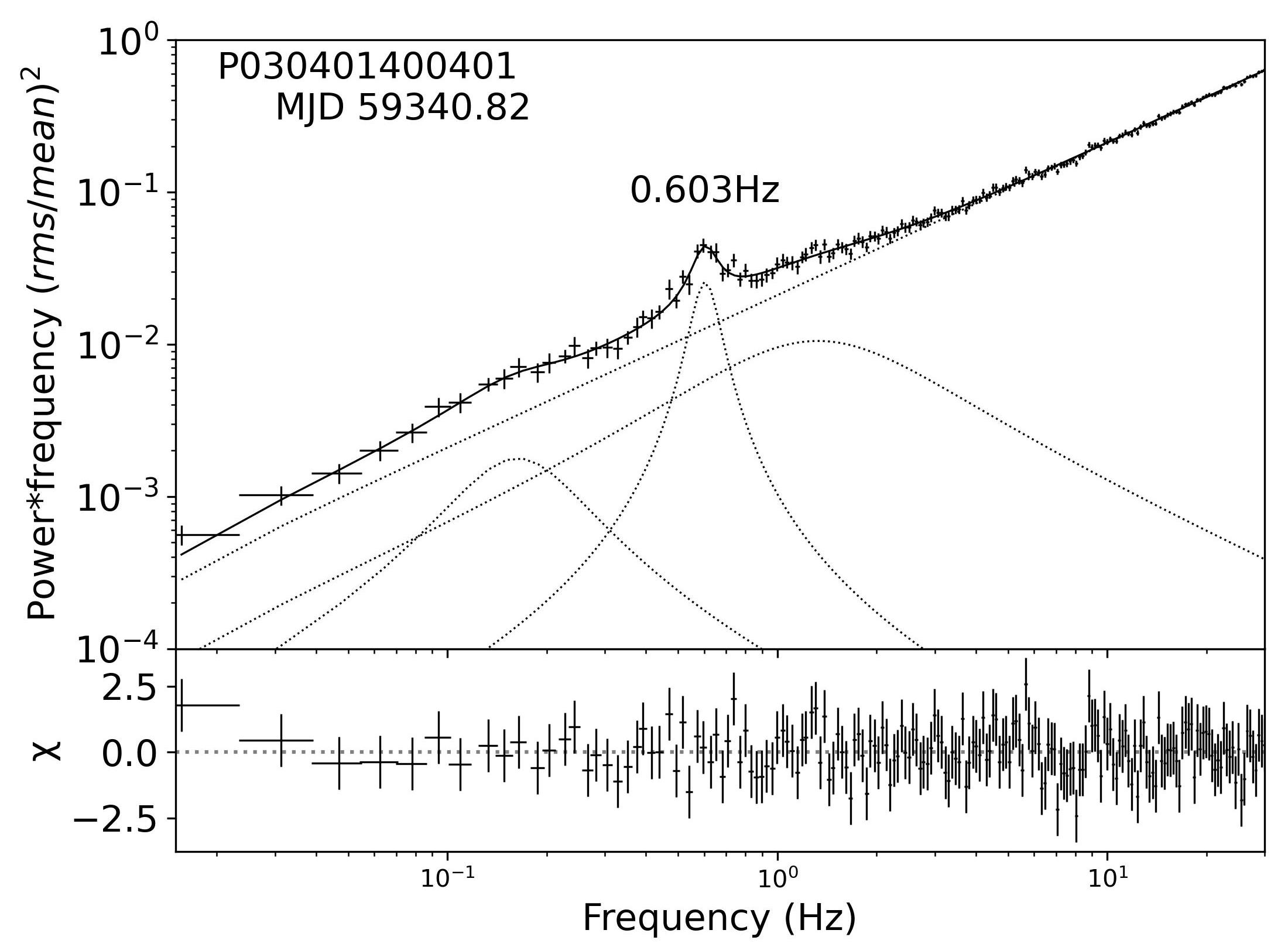}
    \caption{Power density spectra of the observation P030401400201 using the  \textit {Insight}-HXMT LE data (1-10 keV). The solid lines show the best fit with the power$-$law plus multi-Lorentzian function (dotted lines).The ObsID, MJD and QPO fundamental frequency are shown in the panel.}
    \label{fig:QPO}
\end{figure}

\begin{figure}
        \includegraphics[width=\columnwidth]{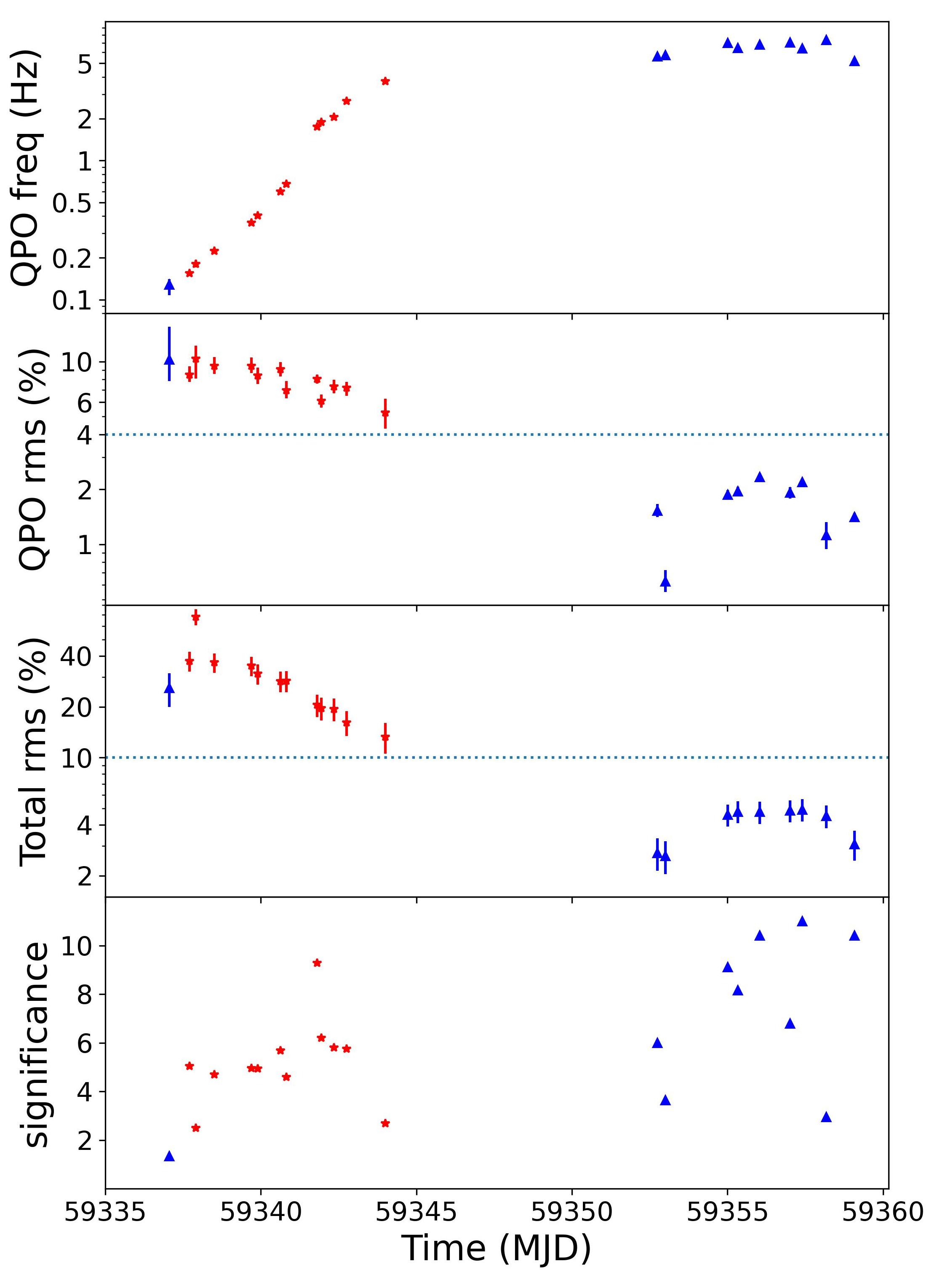}
    \caption{Evolution of the QPO frequencies, the QPO rms and the total rms of MAXI J1803$-$298 in the band of $1-10$ keV as a function of time. The total fractional rms is in 0.01–32 Hz. The significance indexes of QPOs are showed in the bottom panel. The red stars and the blue triangles represent the \textit {Insight}-HXMT and \textit {NICER} data respectively. The error bars of all parameters are presented in Tables 1 and 2, and errors bars of frequencies are smaller than the size of the symbols in the plots. }
    \label{fig:QPO_time}
\end{figure}

\begin{figure*}
        \includegraphics[width=\linewidth]{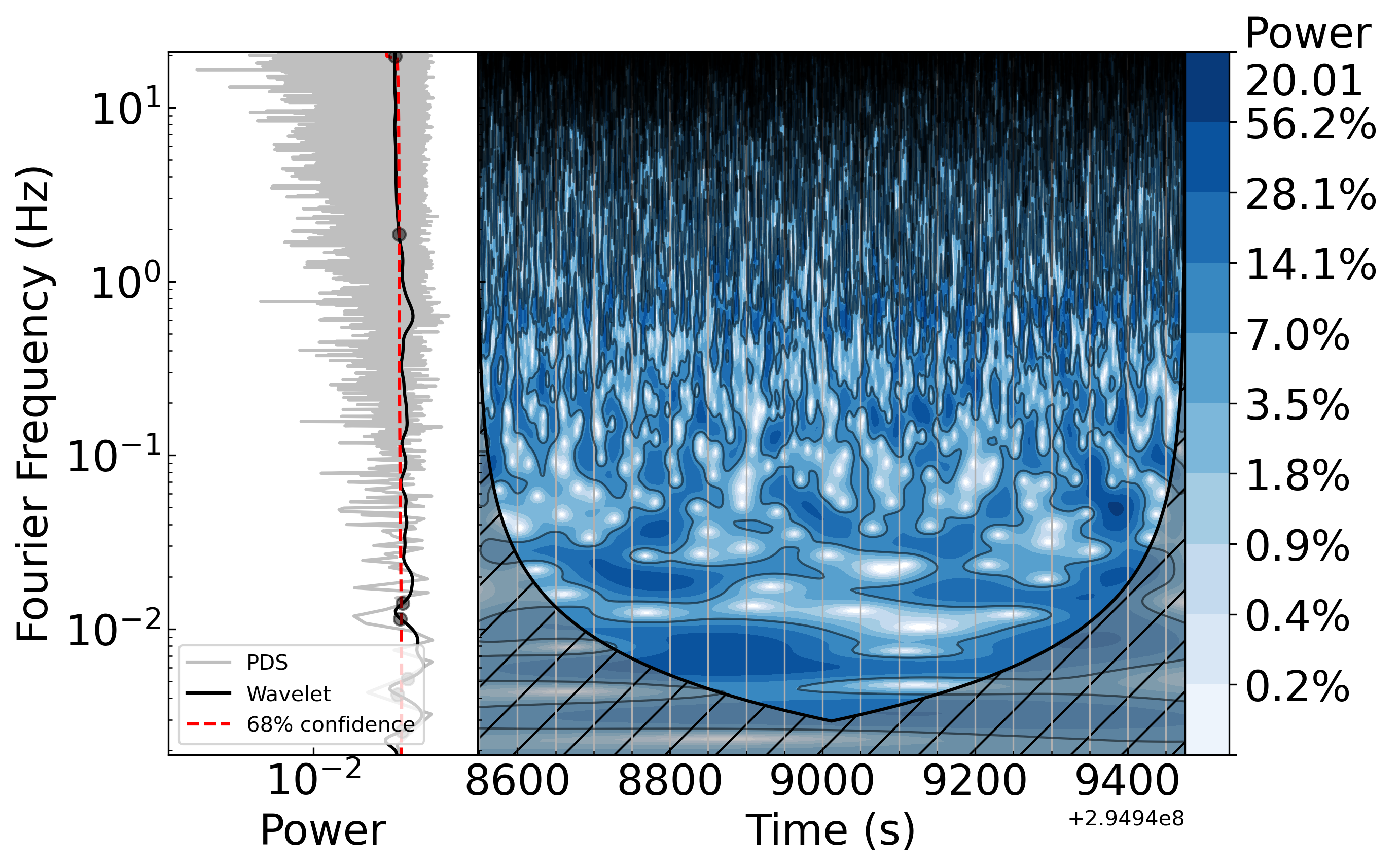}
    \caption{Wavelet result of P030401400401 using the light curve of its longest GTI with 68 \% confidence level. In the right panel, the local wavelet time-frequency spectrum is shown. The black circled areas represent the confidence level of power in the area is greater than 68 \%, and the shaded area represents the cone of influence. The detailed value of power is shown in the color bar on the right side of spectrum. The PDS, global wavelet spectrum and 68 \% confidence spectrum are marked by grey line, black line and red line respectively, in the left panel. }
    \label{fig:complete_wavelet}
\end{figure*}

\begin{figure*}
        \includegraphics[width=\linewidth]{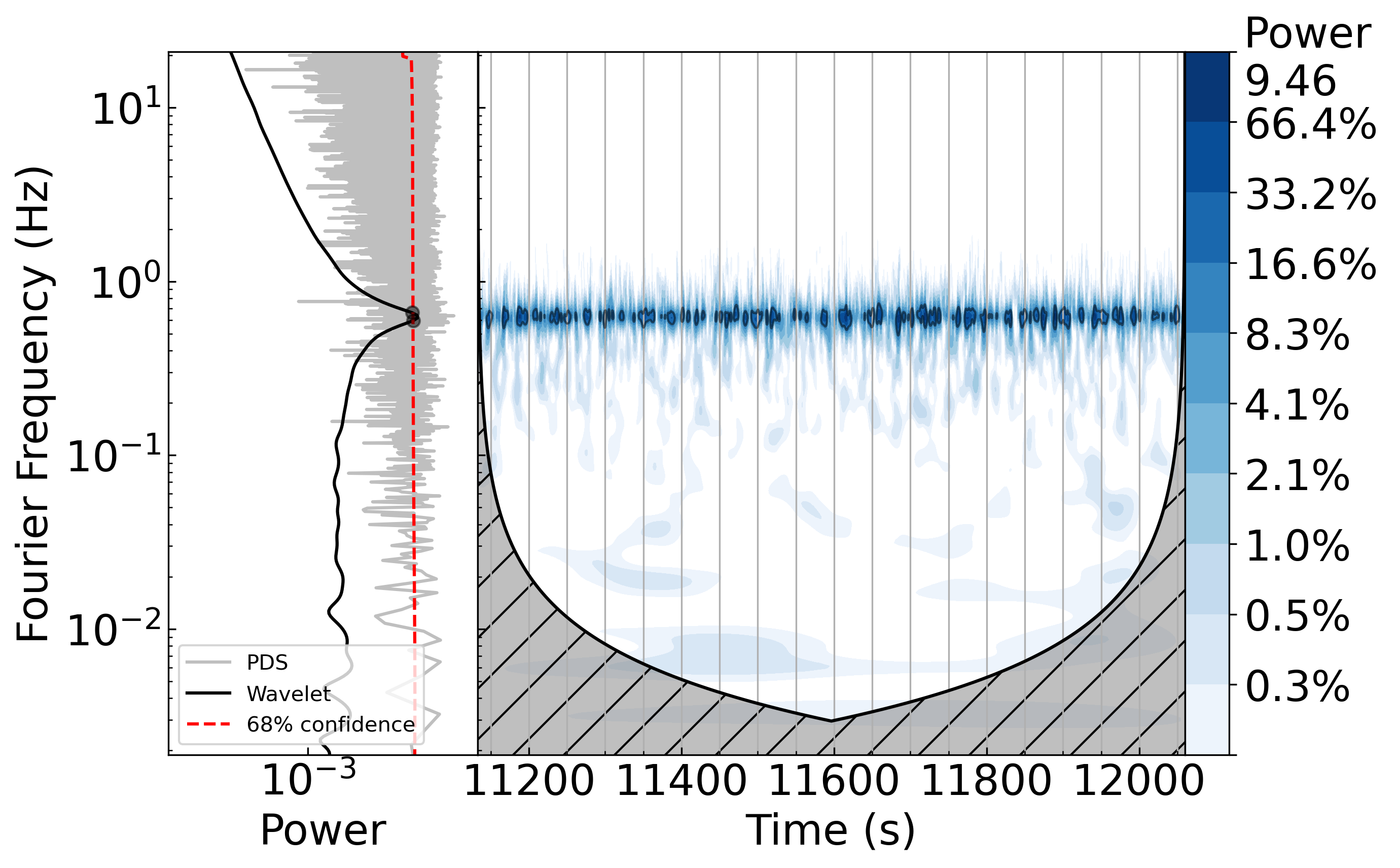}
    \caption{Wavelet result of P030401400401 with QPO component extracted. The selected GTI, confidence level and the elements in the plot are the same as that in Figure~\ref{fig:complete_wavelet}. }
    \label{fig:wavelet}
\end{figure*}

\begin{figure}
        \includegraphics[width=\columnwidth]{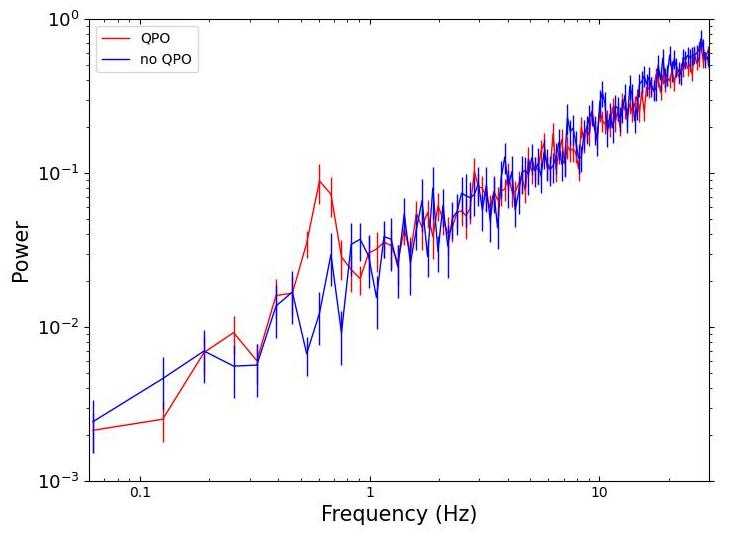}
    \caption{The PDSs of the observation P030401400401 with QPO and no-QPO time intervals are shown with red and blue line, respectively. The time intervals are separated with 68\% confidence level. }
    \label{fig:QPO_separated}
\end{figure}

\begin{figure}
        \includegraphics[width=\columnwidth]{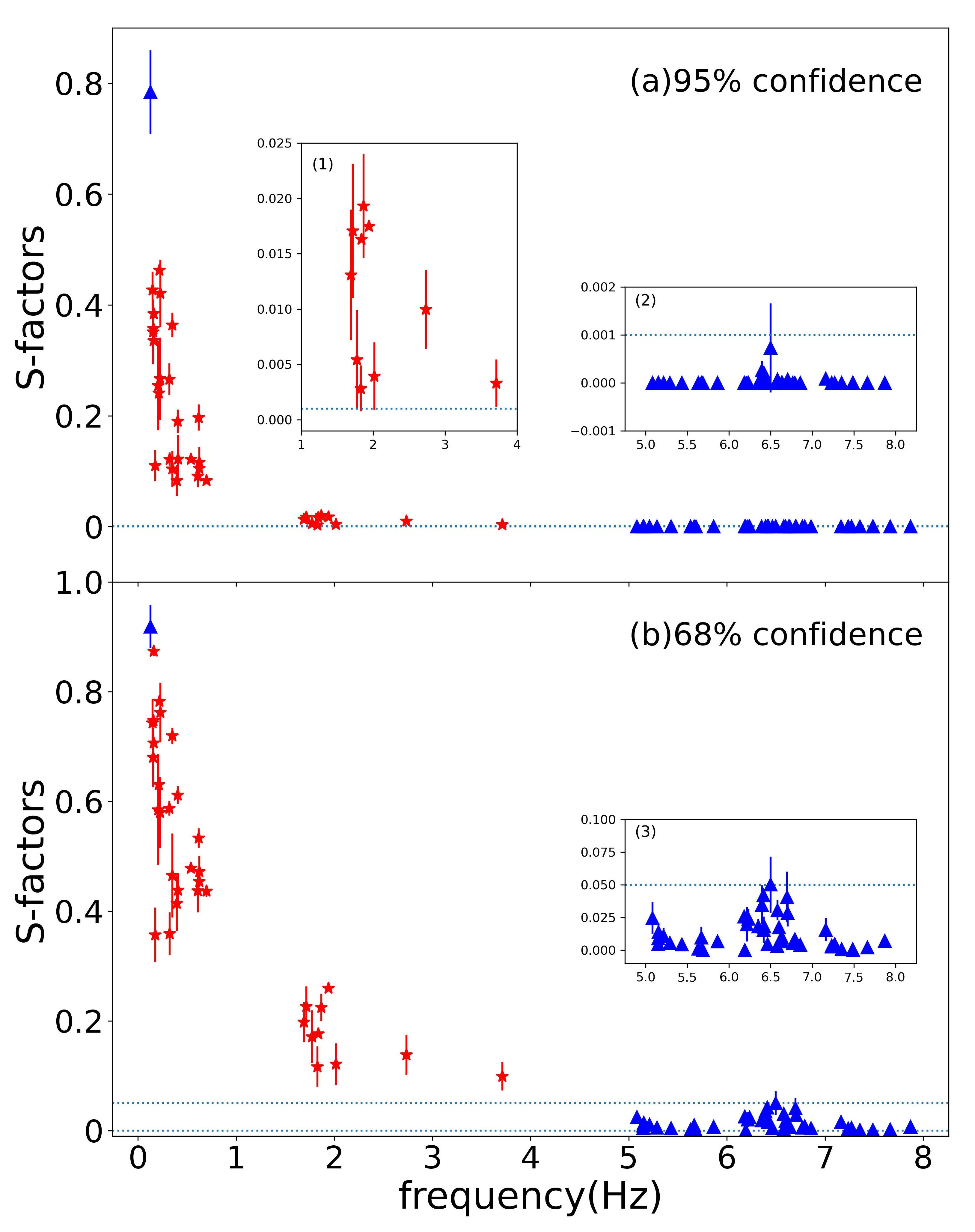}
    \caption{Relation of the S-factors of QPO and QPO frequency in MAXI J1803$-$298 in the band of $1-10$ keV with 95\% (top panel) and 68\% (bottom panel)confidence level. The red stars and the blue triangles represent the \textit {Insight}-HXMT and \textit {NICER} data respectively. The dotted lines mark the range of S-factors of Type$-$B QPOs. The insets in each panel show the points where the S-factor near zero on a smaller Y-axis range. }
    \label{fig:S_freq}
\end{figure}

\begin{figure}
        \includegraphics[width=\columnwidth]{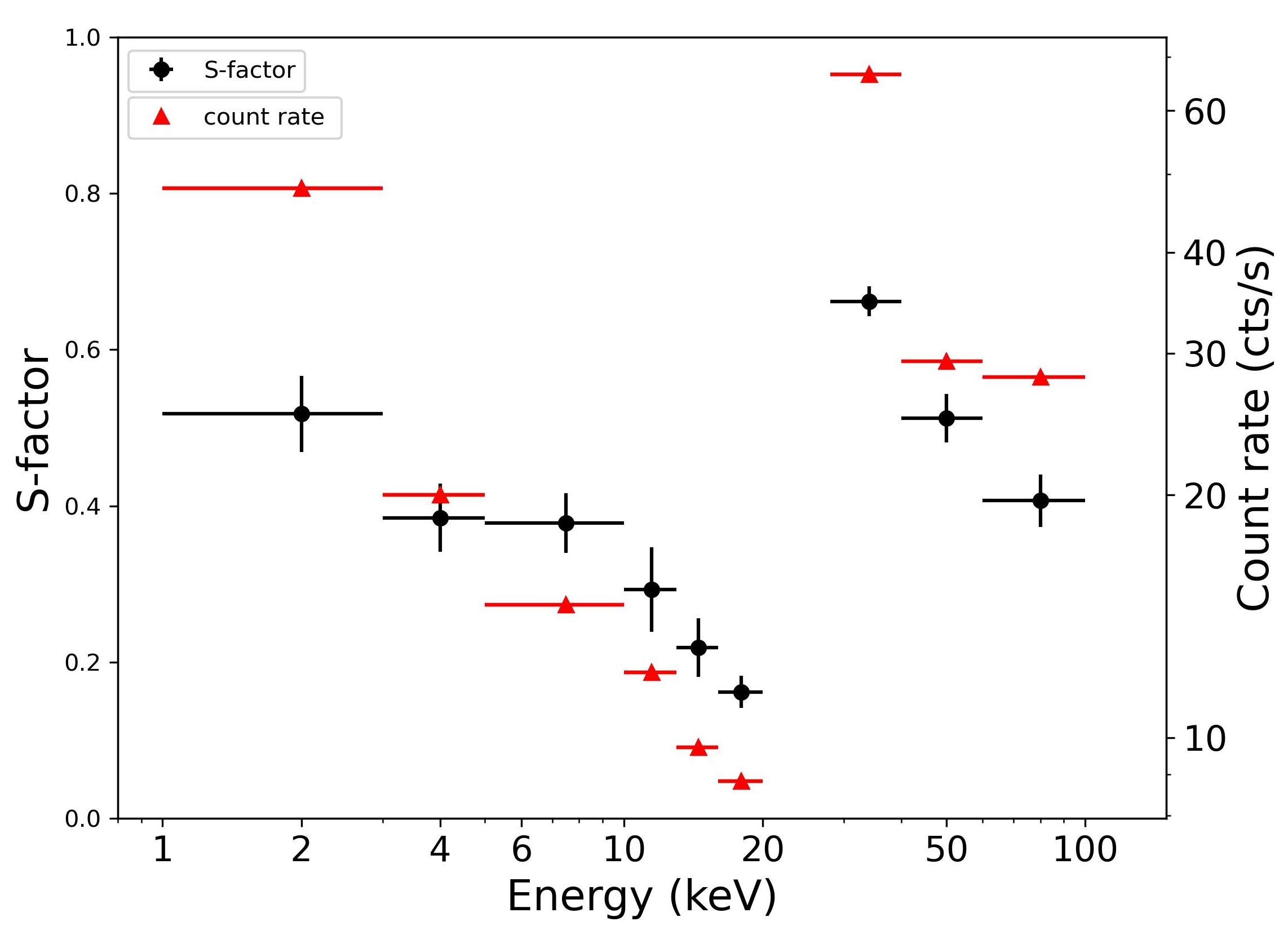}
    \caption{The relation among the photon energy, count rate and S-factor for the observation P030401400101. The black points show the relation between the S$-$factors and energy while the red points show the relation between count and energy. The S-factors are calculated with 68\% confidence level.}
    \label{fig:S_energy}
\end{figure}
\begin{figure*}
        \includegraphics[width=\linewidth]{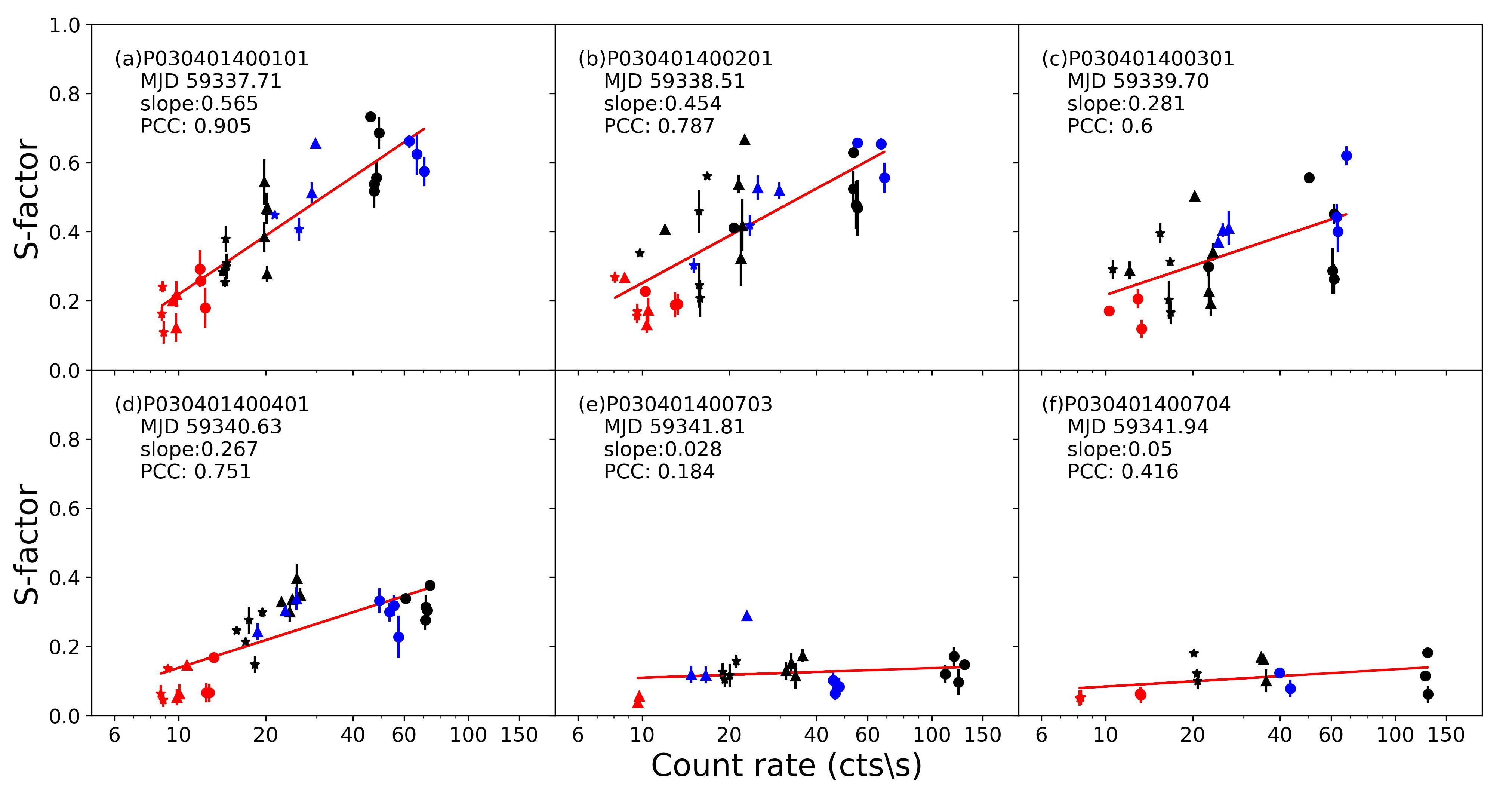}
    \caption{Relation of the S$-$factors of QPOs and count rate for six representative observations with 68\% confidence level. The red lines are the linear regression curves which represent the linear relationship. The ObsID, MJD, the slope of the linear regression curves and the Pearson correlation coefficients (PCCs) of data are shown in the panel.
    The black, red and blue points,triangles and star represent \textit {Insight}-HXMT 1-3, 3-5, 5-10,  10-13, 13-16, 16-20, 28-40, 40-60 and 60-100 keV data, respectively, which are the same as that in Figure~\ref{fig:S_count}. 
    }
    \label{fig:S_count_day}
\end{figure*}

\begin{figure*}
    \centering
    \includegraphics[width=\linewidth]{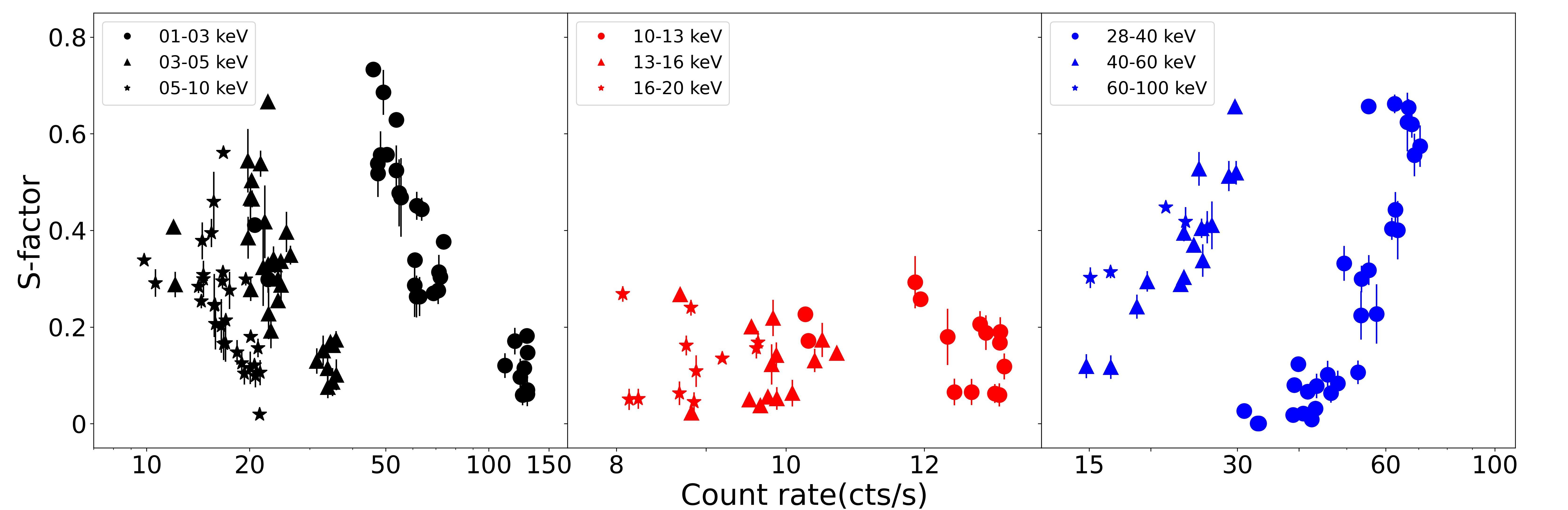}
    \caption{ The relation of S-factor and count rate for different energy band with 68\% confidence level. The black, red, and blue points represent \textit {Insight}-HXMT LE, ME, and HE data respectively.}
    \label{fig:S_count}
\end{figure*}

In Figure~\ref{fig:light curve}, we show the day-averaged light curves of MAXI J1803$-$298 for its 2021 outburst with the \textit {Insight}-HXMT LE ($1-10$ keV) data and the \textit {NICER} ($1-10$ keV) data. Overall, the evolution of count rate for the  \textit {Insight}-HXMT LE ($1-10$ keV) data shows a trend of increasing and then decreasing. The count rate raises from $\sim 100$ cts $s^{-1}$ on MJD 59337 to $\sim 494$ cts $s^{-1}$ on MJD 59348 and then decreases slowly until MJD 59422. An abrupt decrease of count rate around MJD 59355 has been observed in the light curve of both detectors.

The evolution of the hardness ratio with time is also shown in Figure~\ref{fig:light curve}. In this work, the hardness ratio is defined as the count ratio between the energy bands $3-10$ keV and $1-3$ keV. And the error of hardness ratio is calculated based on the error propagation formula. The hardness ratio shows a continuous decreasing trend, from $\sim 0.8$ on MJD 59337 to $\sim 0.2$ on MJD 59430. The hardness intensity diagram (HID) of this outburst has been shown in Figure~\ref{fig:HID}. The x-axis of HID is the hardness ratio while the y-axis is the count rate of $1-10$ keV. Data of each day is plotted as one point. The outburst starts at the hard state (lower right) and moves to the intermediate state (upper left). As seen from the HID, it is obvious that it takes about seven days for outburst to transition from the hard state to the intermediate state. During the hard state, Type-C QPOs have been observed in the \textit {Insight}-HXMT LE ($1-10$ keV) data. Then outburst stays in the intermediate state for several days with high count rate and Type-B QPOs observed in the \textit {NICER} ($1-10$ keV) data. Finally, it reaches the soft state at the lower left of the diagram. 

\subsection{QPO properties}
\label{sec:QPO} 
In 2021 outburst of MAXI J1803$-$298, \cite{Zhu2023} claim that Type-C and Type-B QPOs are observed with \textit {Insight}-HXMT and \textit {NICER}, respectively. We analysis those observations with QPOs and obtain the PDSs. Then we fit them with the model mentioned in Section~\ref{sec:ob}. In Table~\ref{tab:table_hxmt} and Table~\ref{tab:table_ni}, we present a summary of parameters of observations (i.e., Observation ID, MJD  and exposure time) and QPOs (i.e., the centroid frequency $\upsilon_{\rm QPO}$, the FWHM, the QPO and total rms, Q factor and type of QPO) for  \textit {Insight}-HXMT data and  \textit {NICER} data, respectively.

In Figure~\ref{fig:QPO}, we show the PDS of the \textit {Insight}-HXMT observation P030401400401 which has long exposure time and high QPO significance, using its LE ( 1-10 keV) data. The PDSs are fitted with one narrow Lorentzian component which is the typical QPO, two broad Lorentzian components for the broad-band noise and one power-law component with the index fixed at zero for constant white noise.
Figure~\ref{fig:QPO_time} shows the evolution of the QPO centroid frequency, the QPO rms, the total rms and QPO significance as a function of time. It is clear that our data is divided into two stages. The early stage is the LHS and the later stage is the SIMS. The centroid frequency of the QPO quickly increases from 0.1 Hz to 3 Hz in the LHS and stays around 6 Hz in the SIMS. The QPO and total rms both has the obviously decreasing trend with time in LHS but stays stable in SIMS. However, the QPOs rms are all above 4\% in LHS but below in SIMS, while the total rms are all above 10\% in LHS but below in SIMS.  This difference is one of the most important bases for us to classify QPOs.

\subsection{Wavelet analysis}
\label{sec:wavelet} 

A representative wavelet result using the longest GTI of P030401400401 with 68 \% confidence level is shown in Figure~\ref{fig:complete_wavelet}. In the left panel, we show the PDS (grey line), the global wavelet spectrum (black line) and the 68 percent confidence spectrum (red line). The wavelet time-frequency spectrum is shown in the right panel, in which the deeper blue color represent the higher power. The detailed value of power is shown in the color bar on the right side of spectrum. The black circled areas in the spectrum represent the confidence level of power in the circled area is greater than 68 percent, and the shaded area represents the cone of influence (COI). A peak is found around 0.6 Hz in left panel, which is consistent with the frequency of QPO found in the Fourier analysis, thus we consider this peak is the QPO signal. However, some broad-band noise is also recognized as a valid signal, which would affect the wavelet result\citep{Chen2023}. In order to remove the influence of noise, we extract the QPO component from noise with the method mentioned in Section~\ref{sec:ob}. We pick out the QPO component among several Lorentz components, calculate its proportion curve and apply the curve on the wavelet power to obtain the wavelet result with QPO extracted. The wavelet result of same observation with QPO component extracted is shown in the Figure~\ref{fig:wavelet}. 

Without the influence of noise, we can separate the QPO and no-QPO time intervals more accurately for generation of QPO and no-QPO PDSs, as introduced in Section~\ref{sec:ob}. Figure~\ref{fig:QPO_separated} shows the QPO and no-QPO PDSs of P030401400401 with red and blue lines, respectively. An obvious QPO signal appears in the red line but not in the blue line. Furthermore, we use the QPO time interval divided to the total length of GTI to calculate the S-factor as also mentioned in Section~\ref{sec:ob}. Since the QPO time intervals show the time intervals when the power over the red noise background spectrum with a certain confidence level, S-factor describe the distribution of QPOs power, which is similar to the QPO fractional rms\citep{Chen2023}. For comparison, we perform wavelet analysis pipeline with 68\% and 95\% confidence level for all observations to obtain the QPO time interval and calculate the S-factor. Figure~\ref{fig:S_freq} shows the relation of S-factor and QPO frequency with 95\% and 68\% confidence level. As seen from the plots, S-factors decrease with QPO frequency in both two confidence levels. In the top panel (the 95\% confidence level), S-factors quickly decrease with QPO frequency and show few difference between Type-B QPOs and Type-C QPOs with high QPO frequency, and the difference becomes more significant at the 68\% confidence level. S-factors of Type-B QPOs are all below 0.05 and most of them are near zero, while the S-factors of Type-C QPOs are above 0.1 and even larger. 

To study the relation of S-factor and energy dependence of QPOs, we separate the light curve with energy band and perform wavelet analysis for each GTI, as mention in Section~\ref{sec:ob}. Comparing those results, we study the relation of S-factor and count rate in different photon energy. In Figure~\ref{fig:S_energy}, black points represent the S-factors for observation P030401400101 which is the only observation with QPOs clearly discernible in all nine energy bands. The S-factors decrease with energy from 1 keV to 20 keV, and then suddenly increase in ~30 keV and decrease with energy above 50 keV. In addition, we show the count rates for the same observation with red points in Figure~\ref{fig:S_energy}. The count rates also show the correlation of increasing in ~30 keV and decreasing in other energy bands with energy. 

To further study the correlation of the S$-$factors and count rate, for each observation, we calculate the S-factor and count rate of each GTI in different energy bands, as mention in Section~\ref{sec:ob}. In Figure~\ref{fig:S_count_day}, we plot the correlation between S-factor and counts for those observations with more than 10 GTIs. After fitting with linear regression, we obtain the slope of linear regression curve and the Pearson correlation coefficients (PCCs), which are also shown in the panels. It is clear to see that the slope of curves are decreasing with time. Meanwhile, we also study the correlation of S-factor and rate for each energy band, which has been shown in Figure~\ref{fig:S_count}. Different correlations between S-factor and counts are exhibited in different energy bands, with negative correlations in LE data, no apparent correlation in ME data and positive correlations in HE data.

\section{Discussion}
\label{sec:di}

In this work, we mainly study the QPOs detected in the 2021 outburst of MAXI J1803$-$298. The first and foremost part of our study is to distinguish types of QPOs accurately. According to \cite{Ingarm2019}, Type-B QPOs are detected during the SIMS, while Type-C QPOs are more prominent in the LHS and HIMS. The parameters of QPOs, such as the centroid frequency, the Q factor and the rms, and noise related to QPOs are the basis for QPOs classification.

With \textit{Insight}-HXMT data (Table\ref{tab:table_hxmt}), QPOs are observed in the early state of the outburst. During this period, the count rate increases and the hardness decreases, which is the typical LHS. QPOs frequencies increase quickly from $\sim 0.1$ to $\sim 3.5$ Hz in one week with high fractional variability (the QPO rms up to $4\sim10 \%$), which is similar to the Type-C QPOs observed in the previous studies of other sources \citep{Jin2023}. According to \cite{Zhu2023}, Type-B QPOs are detected with NICER data, so we study these ten observations of NICER data (Table\ref{tab:table_ni}). The first observation of \textit{NICER} data (4202130102) is observed at the same day with the first observation of \textit{Insight}-HXMT data, while other observations of \textit{NICER} data are observed after 15$-$22 days, which means observations in the raising stage of the outburst are lacking. Therefore, we can only study this stage of the outburst using the data from \textit{Insight}-HXMT. As seen from Figure~\ref{fig:light curve}, count rate reaches its peak around MJD 59349 and stays stable around 400 cts/s for a few days with a sudden drop around MJD 59355. Hardness also change within a range around 0.5 during the period of last nine observations of NICER data, i.e. MJD 59352 to 59359, which indicates that the source was in the intermediate state.

With the evolution of the QPO parameters plotted in Figure~\ref{fig:QPO_time}, our data can be divided into two different parts. QPOs with an increasing frequency from 0.1 Hz to 3.5 Hz occur before MJD 59345, corresponding to the LHS, while QPOs with a stable frequency around 6 Hz is observed after MJD 59350, corresponding to the intermediate state. 
The QPO rms in LHS is relatively higher (up to ~10\%), which is typical for Type-C QPOs, while the QPO rms values in the intermediate state are all lower than 4\%, indicating that they are Type-B QPOs. The difference in the total rms is also found between two parts. Thus in this work, QPOs with frequency $<$ 4 Hz and $>$ 5 Hz are Type-C QPOs and Type-B QPOs, respectively.

After searching and classifying QPOs in Fourier domain, we extracted QPO components and study them with wavelet transform. For each GTI greater than 192s of each observation, we separate GTI into the QPO and no-QPO time intervals, and calculate the S-factor using the QPO time intervals. The relation of S-factor and QPO frequency has shown in Figure~\ref{fig:S_freq}. For both confidence levels, S-factors have negative correlation with frequency for Type-C QPOs and no specific correlation for Type-B QPOs. With 68\% confidence level, S-factors of Type-B QPOs are all below 0.05 and most of that are near zero, while the S-factors of Type-C QPOs are above 0.1 and even bigger. The negative correlation between S-factor and frequency and the distinction between Type-B QPOs and Type-C QPOs are both consistent with that observed in MAXI J1535$-$571 \citep{Chen2023}.

\cite{Ingram2009} proposed a Type-C QPO model, which assumes a hot, geometrically thick, optically thin inner accretion flow, and cool, geometrically thin, optically thick, truncated outer accretion disc. The inner accretion flow is misaligned with the accretion disc, leading to the Lense-Thirring precession of a radially extended section of the hot inner flow and the rise of Type-C QPOs. As the truncation radius moving inward during the outburst state from the LHS to the HIMS, the radial extent of the hot inner flow becomes smaller causing the higher QPO frequency with the smaller jitters in frequency and the smaller fluctuation power\citep{Ingram2011}. The S-factor describes the distribution of QPO power over time, hence the smaller fluctuation power of inner flow leads to the smaller value of S-factor. The Lense-Thirring precession model provides a possible explanation for the increase of QPO frequency with time and the negative correlation between S-factor and frequency. Besides, the L-T precession model also explains some other observed behaviours of Type-C QPOs, such as the inclination dependence of QPO phase lags \citep{Eijnden2017}. Type-B QPOs may be related to jets \citep{Ruiter2019}, which has much less S-factors, so that possibly with jet formation, the inner corona becomes very small, the frequency is relatively stable around 5 Hz, while the QPO fluctuation power is weak.

For Type-C QPOs with frequency greater than 2 Hz, \cite{Chen2023} found that in MAXI J1535$-$571, which is a high inclination source, the negative correlation between S-factor and frequency is consistent with the statistical results of the rms-frequency correlation in high inclination sources proposed by \cite{Motta2015}, while S-factor and rms show no significant correlations with frequency for Type-B QPOs. \cite{Chen2023} proposed that it makes sense that the S-factor show similar behaviour with rms because they both describe the distribution of QPOs power. Indeed, in our results, the S-factor and rms both show similar negative correlation for Type-C QPOs and no correlation for Type-B QPOs with frequency. However, according to \cite{Motta2015}, in high inclination sources, the rms has opposite correlation for Type-C QPOs with frequency below and above 2 Hz. In our results, for the QPO frequency below 2 Hz, the rms still shows the negative correlation with time, which is inconsistent with the statistical results for high inclination sources but consistent with that for low inclination sources \citep{Motta2015}. However, multiple evidences indicate that MAXI J1803$-$298 is preferred to be a high inclination source \citep{Coughenour2023,Feng2022,Zhu2023}. Hence, either the inclination of MAXI J1803$-$298 needs to be further studied, or different systems may have different thresholds for high inclination as indicated by \cite{Motta2015}.

Thanks to the wide energy range of \textit {Insight}-HXMT, we can study the relation between S-factor of Type-C and the photon energy (black points in Figure~\ref{fig:S_energy}). Negative correlations are found, except for an abrupt increase occurring around 30 keV, which is consistent with the count rate (red points in Figure~\ref{fig:S_energy}) evolution. We consider that changes in the S-factor may correlate with changes in the count rate. Therefore, we plot the relations of S-factor and count rate for six representative observations in Figure~\ref{fig:S_count_day}. After fitting with linear regression, we calculate the slopes of linear regression curves and the Pearson correlation coefficients (PCCs) to measure the linear correlations between S-factors and count rates. The slopes show a decreasing trend with time from $\sim$ 0.6 to $\sim$ 0.05. In addition, the PCCs also show a decreasing trend with time. For observation P030401400101 (left-top panel), the PCC is $\sim$ 0.9, suggesting a near-perfectly positive linear correlation between S-factor and count rate. However, the PCCs become smaller in the MJD 59341 (middle-bottom panel and right-bottom panel), which means that the linear correlations between S-factor and counts become weaker. 

Meanwhile, relations of S-factor and count rate for nine energy bands are plotted in Figure~\ref{fig:S_count}. The correlation of S-factor and counts seems be totally different for \textit {Insight}-HXMT LE, ME, and HE data. We found negative correlations between S-factor and count rate for all three LE bands, but positive correlations for HE bands. For ME bands, S-factor seems to have no correlation with count rate, which is inconsistent with the positive correlation found in MAXI J1535$-$571 with \textit{Insight}-HXMT ME data \citep{Chen2022b}. We speculate that either the low count rate in our ME data or different outburst states leads to this distinction. Nevertheless, the different behaviors of LE and HE data are both consistent with previous studies \citep{Chen2022b,Chen2023}, indicating the different physical origin of Type-C QPOs at high energy and low energy. \cite{Chen2023} propose a possible explanation that this different behaviors of S-factor between LE and HE data may be related to the dual-corona geometry.

The dual-corona model assumes that two coronas are responsible for the variability spectra in BHXRBs \citep{Bellavita2022}. These two coronas have the same power-law photon indices and electron temperatures, but differ in soft-photon source temperatures, the sizes of the coronas, the inner disc radius, and the variability of the external heating rates. The compact and hot corona, with a relatively high effective temperature, is close to the BH, while the large and extended corona, with a lower effective temperature, may be associated with the base of the extended jet. 
Compared with the single-corona model, the dual-corona model have a better fitting for the rms and phase lag spectrum, especially in the relatively low ( < 1.5 keV) and high (> 5 keV) energy bands \citep{Bellavita2022,García2021,Valentina2023}. Therefore, it is reasonable that one of the dual-corona is responsible to the low energy bands, and the other dominates the high energy bands. In addition, there is a phase angle between the QPOs from these two different coronas \citep{García2021}. It is possible that the L-T precession may only happen in the inner corona, the QPOs could propagate to the other corona through the upscattered photons into the flux of seed-photons source \citep{García2021,Bellavita2022}, which may account for a phase angle of two coronas. Due to the different energy dominant regions and dynamics of two coronas, it would also naturally explain the opposite behaviors of S-factors in the LE and HE energy bands. More detailed research on QPOs is still required for studying the mechanism of the dual-corona.

\section{Conclusions}
\label{sec:co}
In this paper, we study the 2021 outburst of MAXI J1803$-$298 with \textit{Insight}-HXMT and \textit{NICER} data. We found QPOs in two different states of this outburst. According to HID, we consider the early state as the LHS and the latter state as the SIMS. With difference on the QPOs rms and the total rms, we confirm that Type-C QPOs with frequency from 0.1 Hz to 3 Hz detected in LHS and Type-B QPOs around 6 Hz detected in SIMS. Base on that, the wavelet analysis is used to study the properties of two types of QPOs. For each observations, the GTIs can be separated into two time regimes: QPOs and no-QPOs. S-factors are calculated, and they decrease with QPO frequency for Type-C QPOs but stay stable around zero for Type-B QPOs. In addition, the correlation of S-factor and counts is determined, and the different correlation of S-factor and count rate for LE and HE data indicates the possibility of different origins of QPOs in high and low energy bands. The new dual-corona model may be a possible explanation for this distinction. More researches about the S-factors of QPOs will be proposed to other black hole X-ray sources in the future.

\section*{Acknowledgements}
We are grateful to the referee for the useful comments and suggestions to improve the paper. This work is supported by the NSFC (No. 12233006, 12133007 and 12373046), the National Key Research and Development Program of China (Grants No. 2021YFA0718503) and the Foundations from Yunnan Province (No. 202301AS070073).

\section*{Data Availability}
Data used in this work are from the Institute of High Energy Physics, Chinese Academy of Sciences (IHEP-CAS) and have been publicly available for download from the \textit{Insight}-HXMT website http://hxmtweb.ihep.ac.cn/.



\bibliographystyle{mnras}
\bibliography{example} 




\appendix
\begin{table*}
    \centering
    \caption{Low-frequency QPO parameters for MAXI J1803$-$298 with the \textit{Insight}-HXMT LE (1-10 keV) data.}
    \label{tab:table_hxmt}
    \begin{tabular}{cccccccccccc}
    \hline
    \hline
        No. &Observation ID & MJD & Exposure & $\nu_{\rm QPO}$ & FWHM & Q & the QPO rms & the total rms & significance & type \\
        &&&(s)&(Hz)&(Hz)&&(\%)&(\%)\\
    \hline
        \\1 & P030401400101 & 59337.71 & 3941 & $0.155_{-0.002}^{+0.003}$ & $0.027_{-0.008}^{+0.017}$ & $5.59\pm 0.48$ & $8.52_{-0.761}^{+0.922}$ & $37.34\pm 4.96$ & 5.01&C\\
        \\2 & P030401400102 & 59337.91 & 1320 & $0.180_{-0.09}^{+0.012}$ & $0.050_{-0.027}^{+0.020}$ & $3.60\pm 0.23$ & $10.37_{-2.302}^{+1.869}$ & $68.12\pm 7.43$ & 2.48&C\\
        \\3 & P030401400201 & 59338.51 & 2940 & $0.223_{-0.005}^{+0.005}$ & $0.058_{-0.014}^{+0.019}$ & $3.82\pm 0.29$ & $9.47_{-0.889}^{+1.131}$ & $36.62\pm 4.83$ & 4.68&C\\
        \\4 & P030401400301 & 59339.70 & 3229 & $0.358_{-0.007}^{+0.008}$ & $0.105_{-0.020}^{+0.023}$ & $3.38\pm 0.21$ & $9.47_{-0.866}^{+1.080}$ & $34.96\pm 4.58$ & 4.95&C\\
        \\5 & P030401400302 & 59339.89 & 1200 & $0.403_{-0.010}^{+0.010}$ & $0.063_{-0.022}^{+0.024}$ & $6.32\pm 0.36$ & $8.39_{-0.838}^{+0.860}$ & $31.38\pm 4.30$ & 4.94&C\\
        \\6 & P030401400401 & 59340.63 & 3147 & $0.603_{-0.007}^{+0.007}$ & $0.104_{-0.024}^{+0.025}$ & $4.80\pm 0.40$ & $8.27_{-0.764}^{+0.835}$ & $28.37\pm 3.94$ & 7.55&C\\
        \\7 & P030401400402 & 59340.82 & 1140 & $0.680_{-0.007}^{+0.008}$ & $0.049_{-0.014}^{+0.035}$ & $13.72\pm 0.50$ & $6.94_{-0.645}^{+0.867}$ & $28.43\pm 4.02$ & 4.59&C\\
        \\8 & P030401400703 & 59341.81 & 2400 & $1.747_{-0.015}^{+0.015}$ & $0.286_{-0.042}^{+0.048}$ & $6.09\pm 0.15$ & $8.02_{-0.429}^{+0.435}$ & $20.50\pm 3.06$ & 9.27&C\\
        \\9 & P030401400704 & 59341.94 & 1860 & $1.885_{-0.017}^{+0.017}$ & $0.181_{-0.033}^{+0.037}$ & $10.39\pm 0.19$ & $6.09_{-0.048}^{+0.050}$ & $19.60\pm 2.97$ & 6.19&C\\
        \\10 & P030401400707 & 59342.35 & 1140 & $2.049_{-0.029}^{+0.028}$& $0.323_{-0.064}^{+0.080}$ & $6.34\pm 0.22$ & $7.31_{-0.618}^{+0.642}$ & $19.37\pm 2.99$ & 5.79&C\\
        \\11 & P030401400710 & 59342.75 & 956 & $2.671_{-0.050}^{+0.050}$ & $0.503_{-0.101}^{+0.105}$ & $5.31\pm 0.20$ & $7.18_{-0.683}^{+0.564}$ & $16.14\pm 2.72$ & 5.75&C\\
        \\12 & P030401400804 & 59344.00 & 240 & $3.711_{-0.058}^{+0.062}$ & $0.225_{-0.105}^{+0.186}$ & $16.42\pm 0.64$ & $5.24_{-0.953}^{+0.999}$ & $13.30\pm 2.73$ & 2.68&C\\
    \hline
    \end{tabular}
\end{table*}

\begin{table*}
    \centering
    \caption{Low-frequency QPO parameters for MAXI J1803$-$298 with the NICER (1-10 keV) data.}
    \label{tab:table_ni}
    \begin{tabular}{cccccccccccc}
    \hline
    \hline
        No. &Observation ID & MJD & Exposure & $\nu_{\rm QPO}$ & FWHM & Q & the QPO rms & the total rms & significance & type \\
        &&&(s)&(Hz)&(Hz)&&(\%)&(\%)\\
    \hline
        \\1 & 4202130102 & 59337.05 & 536 & $0.128_{-0.020}^{+0.002}$ & $0.037_{-0.023}^{+0.066}$ & $3.41\pm 1.20$ & $10.28_{-2.447}^{+5.269}$ & $25.75\pm 5.78$ & 1.33& C\\
        \\2 & 4202130105 &59352.74 & 2680 & $5.603_{-0.118}^{+0.135}$ & $1.654_{-0.358}^{+0.474}$ & $3.38\pm 0.25$ & $1.53_{-0.122}^{+0.133}$ & $2.73\pm 0.59$ & 5.98& B\\
        \\3 & 4202130106 &59353.00 & 12271 & $5.718_{-0.061}^{+0.064}$ & $0.552_{-0.166}^{+0.226}$ & $10.35\pm 0.35$ & $0.62_{-0.075}^{+0.096}$ & $2.62\pm 0.58$ & 3.63& B\\
        \\4 & 4675020101 &59355.01 & 5818 & $7.019_{-0.040}^{+0.039}$ & $1.002_{-0.117}^{+0.124}$ & $7.00\pm 0.12$ & $1.87_{-0.101}^{+0.104}$ & $4.59\pm 0.68$ & 9.11& B\\
        \\5 & 4202130108 &59355.33 & 3819 & $6.432_{-0.027}^{+0.026}$ & $0.669_{-0.099}^{+0.104}$ & $9.61\pm 0.15$ & $1.94_{-0.117}^{+0.121}$ & $4.79\pm 0.71$ & 8.15& B\\
        \\6 & 4675020102 &59356.04 & 3012 & $6.816_{-0.050}^{+0.048}$ & $1.193_{-0.156}^{+0.171}$ & $5.71\pm 0.13$ & $2.33_{-0.114}^{+0.109}$ & $4.77\pm 0.72$ & 10.41& B\\
        \\7 & 4675020103 &59357.01 & 5381 & $7.037_{-0.067}^{+0.067}$ & $1.293_{-0.170}^{+0.182}$ & $5.44\pm 0.13$ & $1.91_{-0.136}^{+0.145}$ & $4.86\pm 0.72$ & 6.78& B\\
        \\8 & 4202130110 &59357.40 & 7627 & $6.383_{-0.025}^{+0.025}$ & $0.889_{-0.086}^{+0.090}$ & $7.17\pm 0.09$ & $2.19_{-0.101}^{+0.097}$ & $4.93\pm 0.74$ & 11.00& B\\
        \\9 & 4675020104 &59358.17 & 2762 & $7.346_{-0.093}^{+0.100}$ & $0.764_{-0.275}^{+0.341}$ & $9.61\pm 0.40$ & $1.12_{-0.176}^{+0.203}$ & $4.51\pm 0.68$ & 2.94& B\\
        \\10 & 4202130112 &59359.08 & 4599 & $5.179_{-0.032}^{+0.031}$ & $0.580_{-0.067}^{+0.079}$ & $8.92\pm 0.12$ & $1.41_{-0.065}^{+0.070}$ & $3.07\pm 0.62$ & 10.40& B\\
    \hline
    \end{tabular}
\end{table*}

\end{document}